%% file: main.tex
\documentclass[runningheads]{llncs}
\usepackage[T1]{fontenc}
\usepackage{amsmath,amssymb,amsfonts}
\usepackage{algpseudocode}
\usepackage[linesnumbered,ruled,vlined]{algorithm2e}
\usepackage{pifont}
\usepackage{graphicx}
\usepackage{textcomp}
\usepackage{xcolor}
\usepackage{pgf}
\usepackage{tikz}
\usepackage{color}
\usepackage{stfloats}
\usepackage{multirow}
\usepackage{diagbox}
\usepackage[hyphens]{url}
\usepackage{parallel}
\usepackage{wrapfig}
\usepackage{marvosym}
\usepackage[normalem]{ulem}
\usepackage[sort,nocompress]{cite}

\usepackage{booktabs}

\usepackage{comment}
\usepackage[bookmarks=true,breaklinks=true,colorlinks,linkcolor=black,citecolor=blue,urlcolor=black]{hyperref}

\newcommand{\squishlist}{
   \begin{list}{$\bullet$}
    { \setlength{\itemsep}{0pt}      \setlength{\parsep}{0pt}
      \setlength{\topsep}{3pt}       \setlength{\partopsep}{0pt}
      \setlength{\listparindent}{-2pt}
      \setlength{\itemindent}{-5pt}
      \setlength{\leftmargin}{1em} \setlength{\labelwidth}{0em}
      \setlength{\labelsep}{0.5em} } }

\newcommand{\squishend}{
    \end{list}  }

\newcommand*\blackcircledempty[1]{\tikz[baseline=(char.base)]{
        \node[shape=circle, text={rgb,255:red,0;green,0;blue,0}, font=\small, draw={rgb,255:red,0;green,0;blue,0},inner sep=0.5pt] (char) {#1};}}

\newcommand{\revise}[1]{{\color{blue}#1}}

\begin{document}
\title{ADE-HGNN: Accelerating HGNNs through Attention Disparity Exploitation}
%
%
\author{Dengke Han\inst{1,2}, Meng Wu\inst{1}, Runzhen Xue\inst{1,2}, \\
Mingyu Yan\inst{1}\textsuperscript{(\Letter)}, Xiaochun Ye\inst{1}, and Dongrui Fan\inst{1}}
%
\authorrunning{D. Han et al.}
%

\institute{SKLP, Institute of Computing Technology, Chinese Academy of Sciences \and
University of Chinese Academy of Sciences \\
\email{\{handengke21s, wumeng, xuerunzhen21s, yanmingyu, yexiaochun, fandr\}@ict.ac.cn}}

\maketitle              
\input{abstract}
\input{introduction}

\input{background}

\input{motivation}
\input{design}

\input{experiments}
\input{related_work}
\input{conclusion}
%
%
%
\bibliographystyle{splncs04}
\bibliography{refs}
%




\end{document}

%% file: abstract.tex
\begin{abstract}
Heterogeneous Graph Neural Networks (HGNNs) have recently demonstrated great power in handling heterogeneous graph data, rendering them widely applied in many critical real-world domains. Most HGNN models leverage attention mechanisms to significantly improve model accuracy, albeit at the cost of increased computational complexity and memory bandwidth requirements. Fortunately, the attention disparity from source vertices towards a common target vertex unveils an opportunity to boost the model execution performance by pruning unimportant source vertices during neighbor aggregation. 

In this study, we commence with a quantitative analysis of the attention disparity in HGNN models, where the importance of different source vertices varies for the same target vertex. To fully exploit this finding for inference acceleration, we propose a runtime pruning method based on min-heap and map it to a dedicated hardware pruner to discard unimportant vertices. Given that the pruning overhead itself is non-negligible and cannot be amortized by conventional staged execution paradigm, an operation-fusion execution flow of HGNNs is introduced to overlap the pruning overhead while harnessing inter-stage parallelism. Finally, we present the design of a novel HGNN accelerator, ADE-HGNN, tailored to support the proposed execution framework. Our experimental results demonstrate that ADE-HGNN achieves an average performance improvement of 28.21$\times$ over the NVIDIA GPU T4 platform and 7.98$\times$ over the advanced GPU A100, with the inference accuracy loss kept within a negligible range of 0.11\%$\sim$1.47\%. Furthermore, ADE-HGNN significantly reduces energy consumption to 1.97\% and 5.37\% of the two platforms, respectively.

\keywords{Heterogeneous Graph Neural Network  \and HGNN Accelerator \and Vertex Pruning \and Operation Fusion}
\end{abstract}

%% file: introduction.tex
\section{Introduction}

Owing to their remarkable power in processing non-Euclidean graph data, Graph Neural Networks (GNNs) have been widely applied across various critical domains. The early success of GNNs has primarily revolved around homogeneous graphs (HomoGs) consisting of a single type of vertices and edges. However, many real-world data in complex systems are more aptly represented as heterogeneous graphs (HetGs). In contrast to HomoGs, HetGs encompass multiple types of vertices and edges, embodying not only the structural information inherent in graph data but also the semantic information harbored in the relations. Due to the powerful representation ability of HetGs, recently, Heterogeneous Graph Neural Networks (HGNNs) have been adopted in many critical fields including recommendation systems~\cite{recommendation_systems}, cybersecurity~\cite{cybersecurity}, and many others.

The execution process of mainstream HGNNs typically comprises four primary stages~\cite{R-GAT,HAN,Simple-HGN}: \textit{Semantic Graph Build (SGB)}, \textit{Feature Projection (FP)}, \textit{Neighbor Aggregation (NA)} and \textit{Semantic Fusion (SF)}. During the whole execution process, NA stands as the predominant stage\cite{understand_HGNN}. The majority of mainstream HGNN models leverage attention mechanisms in the NA stage~\cite{HAN, MAGNN, R-GAT, Simple-HGN} to enhance the model accuracy. However, the introduction of attention mechanisms increases the computational complexity and exacerbates random memory access characteristics, resulting in decreased model execution efficiency. Fortunately, the attention mechanism assigns varying levels of importance to different source vertices concerning the same target vertex. Through selective elimination of relatively unimportant neighbor vertices, the performance of model execution can be dramatically improved with negligible inference accuracy loss.

Due to the recent emergence of HGNNs, there are only very few dedicated acceleration efforts for them. Existing work~\cite{MetaNMP} has not yet explored the potential benefits of leveraging attention disparity to prune and accelerate HGNN inference execution. Moreover, pruning neighbors on conventional platforms faces significant challenges due to the limitations of general-purpose hardware structures and staged execution paradigms. In this work, we first propose a runtime neighbor pruning method based on min-heap to efficiently discard unimportant neighbors. Then we introduce an operation-fusion-based execution flow to amortize the pruning overhead as well as harness the inter-stage parallelism. Our proposed HGNN accelerator, ADE-HGNN, achieves significant performance improvements and energy savings compared to traditional GPU platforms, with negligible loss in inference accuracy. We summarize our contributions as follows:
\par
\squishlist
\item
We conduct a quantitative analysis of the attention disparity in HGNN models and highlight the substantial challenges of utilizing this opportunity for acceleration on traditional GPU platforms.
\item
To leverage attention disparity for inference acceleration, we propose a runtime neighbor pruning method based on min-heap and a novel operation-fusion HGNN execution flow to amortize the additional overhead induced by pruning.
\item 
We design a novel HGNN accelerator, ADE-HGNN, equipped with a unified computing unit to support the optimized execution process of HGNN models.
\item
We conduct comprehensive experiments to demonstrate the advantages of our work. The experimental results indicate that ADE-HGNN achieves an average speedup of 28.21$\times$ over NVIDIA GPU T4 and 7.98$\times$ over the advanced GPU A100, with a negligible inference accuracy loss within range of 0.11\%$\sim$1.47\%. Additionally, the energy consumptions are reduced to only 1.97\%  and 5.37\%.
\squishend

%% file: background.tex
\section{Background}
\subsection{Heterogeneous Graph and Semantic Graph}

\begin{figure*}[!ht] 
	\centering
	\includegraphics[width=1\textwidth]{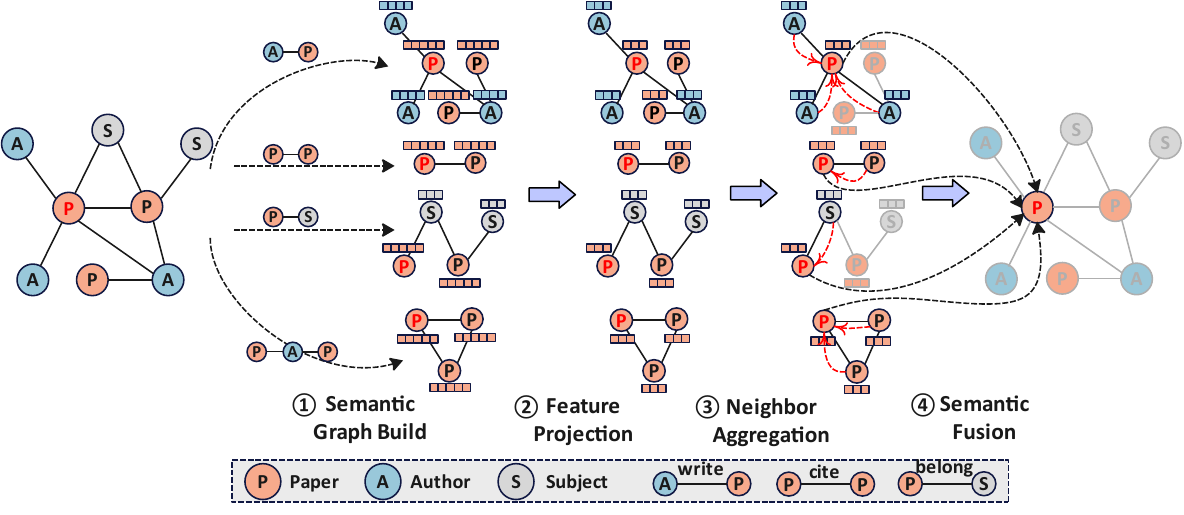}
	\caption{An example of HetGs and execution process of HGNN models.}
        \vspace{-24pt}
	\label{fig:HGNN}
\end{figure*}

In contrast to HomoGs, HetGs encompass multiple types of vertices and relations, embodying both structural and semantic information. Fig.~\ref{fig:HGNN} illustrates a simple example of a HetG from the ACM dataset, including three types of vertices along with three types of adjacency relations between them, which are abbreviated as AP, PP and PS. In addition to simple relations like AP, different combinations of these relations constitute higher-order relations like PAP, referred to as metapaths. Each type of relation or metapath represents unique semantic information between the two endpoints connected. The semantic graphs are derived from the original heterogeneous graph based on specific relations or metapaths, with each semantic graph containing only one type of them. A part of HGNN models~\cite{HAN,MAGNN} utilizes metapaths to construct semantic graphs, while others~\cite{R-GCN,R-GAT} employ simple relations, i.e., edge types, for semantic graph construction.

\subsection{Heterogeneous Graph Neural Networks}

To capture both the structural information and semantic information in HetGs, most prevalent HGNN models contain four primary execution stages as shown in Fig.~\ref{fig:HGNN}.
\blackcircledempty{1} \textit{Semantic Graph Build}: The SGB stage builds semantic graphs for the following stages by partitioning the original HetG into several semantic graphs based on relations or predefined metapaths.
\blackcircledempty{2} \textit{Feature Projection}: The original feature vectors are transformed into new ones using an MLP within each semantic graph during FP stage.
\blackcircledempty{3} \textit{Neighbor Aggregation}: The NA stage extracts structural information by aggregating neighboring features within the same semantic graph.
\blackcircledempty{4} \textit{Semantic Fusion}: The SF stage combines the results of the NA stage across different semantic graphs for each target vertex to gain semantic information.

NA is considered the predominant stage among all stages above~\cite{understand_HGNN}. The majority of mainstream HGNN models leverage weighted aggregation in NA stage by introducing an attention mechanism~\cite{GAT} to encourage the model to focus on relatively more important neighbors. The attention importance is calculated using the vertex features of both the source and target vertices, which are represented as:

\begin{equation}
\begin{gathered}
\theta_{uv}={\rm LeakyReLU}(a^T[W^{c_u}\cdot h_u || W^{c_v}\cdot h_v])={\rm LeakyReLU}(a^T[h_u' || h_v']), \\
\alpha_{uv}=Softmax(\theta_{uv})=\frac{\exp{(\theta_{uv})}}{\sum_{t\in \mathcal{N}_v} \exp{(\theta_{tv})}}.
\label{eq:attn}
\end{gathered}
\end{equation}

$h$ represents the original feature vector of vertices, while $W^c$ denotes the type-specific projection matrix. By performing the dot product with a pre-trained learnable attention vector $a^T$, the attention coefficient $\theta_{uv}$ of edge $e_{uv}$ is computed from the concatenated projected features of vertices $u$ and $v$. After obtaining the attention importance $\alpha_{uv}$ through $Softmax$, the attention weighted aggregation is performed as $h_v = \sum_{u\in {\mathcal{N}_v\bigcup{\{v\}}}} \alpha_{uv}\cdot{h_u'}$, where $\mathcal{N}_v$ denotes the neighboring set of vertex $v$, and $h_u'$ symbolizes the projected feature of vertex $u$.

%% file: motivation.tex
\section{Motivation}

\subsection{Attention Disparity}

\begin{figure}[!htb] 
	\centering
	\vspace{-12pt}
	\includegraphics[width=1\textwidth]{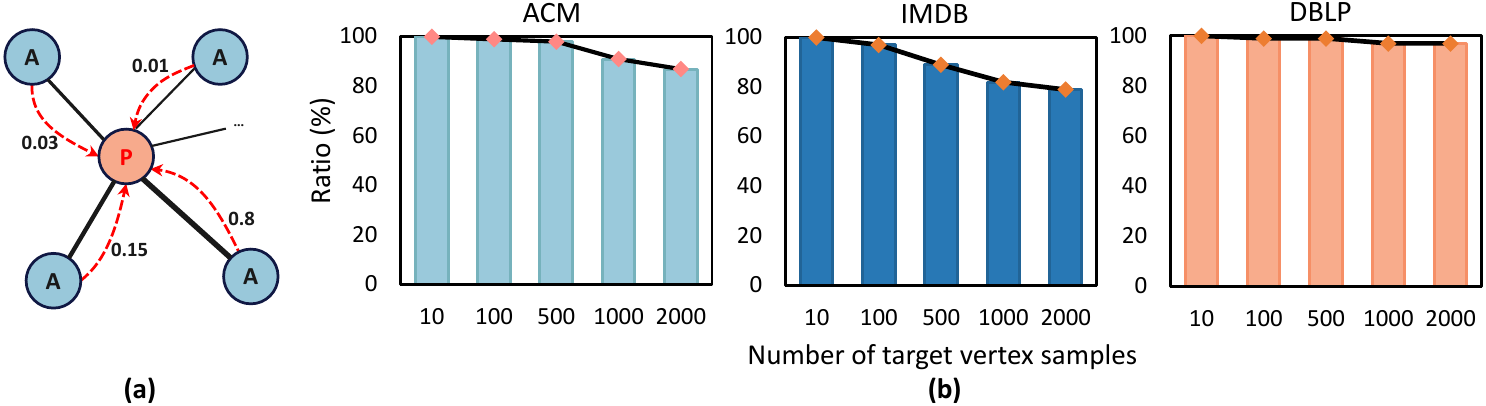}
	\caption{The attention disparity: (a) Varying attention importance; (b) Average ratio of the accumulated attention importance of the top 20\% neighbors.}
	\label{fig:attention_distribution}
\end{figure}

Attention disparity refers to the phenomenon where neighboring vertices exhibit varying attention importance towards a shared target vertex, as shown in Fig.~\ref{fig:attention_distribution}(a). We conduct experiments employing the HAN model\cite{HAN} across three prevalent datasets, randomly sampling a set of target vertices and retaining only the top 20\% of their neighbors with the largest attention importance. Fig.~\ref{fig:attention_distribution}(b) illustrates the average ratio of importance for the selected neighbors compared to the total attention importance of all neighbors, i.e., $ratio=({\sum_{v\in{V_{sample}}}({\sum_{u\in\hat{\mathcal{N}_{v}} }{\alpha_{uv}}/\sum_{k\in\mathcal{N}_v}{\alpha_{kv})}}})/{\left|V_{sample}\right|}$, where $V_{sample}$ is the set of randomly sampled target vertices, $\hat{\mathcal{N}_{v}}$ represents the top 20\% neighbors of $v$. Notably, a minority of neighboring vertices contribute significantly more attention importance than others, indicating substantial information redundancy. While this proportion decreases with an increasing number of sampled target vertices due to the bias of data distribution induced by randomness, it remains relatively high, with a worst-case average of 87.36\%. This phenomenon highlights that a selected subset of crucial vertices are able to provide the necessary information for downstream tasks already. By leveraging this insight, the HGNN model's execution can be significantly expedited by reducing the computational workload and off-chip memory access through pruning unimportant vertices.

\subsection{Challenge to Exploit the Opportunity}
\label{sec:challenge}

\begin{figure}[!ht] 
	\centering
	\includegraphics[width=0.88\textwidth]{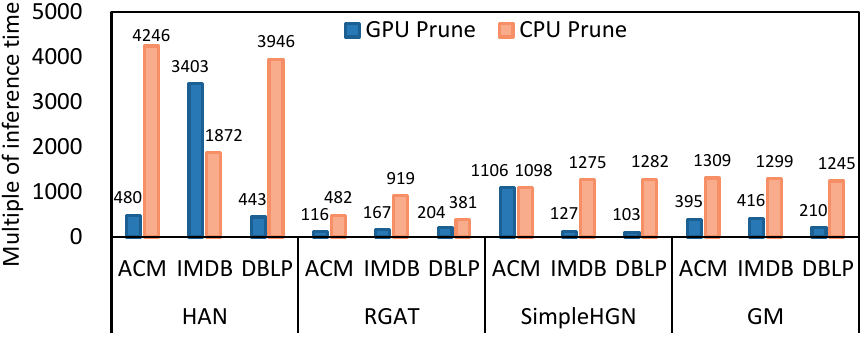}
	\caption{The ratio of pruning time on GPU and CPU to inference time on GPU.}
	\label{fig:pruning_overhead}
\end{figure}

When executing the HGNN models on traditional GPU platforms, a staged execution paradigm is typically employed, following the FP-NA-SF sequence~\cite{understand_HGNN}. In this scenario, pruning vertices based on attention importance necessitates sorting neighboring vertices by their importance, along with a complex control flow involving neighbor extraction and edge re-indexing, which incurs significant overhead and results in a dramatic degradation in overall performance. Fig.~\ref{fig:pruning_overhead} reveals that when performing HGNN inference on an NVIDIA GPU T4 with an Intel Xeon Silver 4208 CPU, the overhead introduced by neighbor pruning based on attention importance greatly exceeds the inference time itself. On a geometrically average, pruning on the GPU takes 325.91$\times$ longer than the inference time, while it's 1284.13$\times$ on the CPU, which is deemed unacceptable. 

Accordingly, there exists a significant challenge in effectively leveraging the attention disparity for neighbor pruning on a traditional platform. This is due to the unsuitability of GPUs for sorting and intricate control operations compared to parallel computing and the failure of the staged execution approach to amortize the pruning overhead, while CPUs inherently exhibit inefficient execution. Consequently, a novel accelerator with an optimized execution flow is required to fully exploit the acceleration potential offered by attention disparity.

%% file: design.tex
\section{Optimized HGNN Execution Flow}

\subsection{Decomposition of Attention Computation}
As illustrated in Eq.~\ref{eq:attn}, the attention importance of a source vertex $u$ to a target vertex $v$ is computed from the projected features of $u$ and $v$. For each edge, the features of the two endpoints have to be randomly accessed, and the attention coefficients for each of these two endpoints need to be recalculated again. To reduce random memory accesses and eliminate unnecessary redundant computations, we transform this process into the following equivalent formula:

\begin{align}
\label{eq:coefficient_transform}
\theta_{uv} & = LeakyReLU(a^{T}\cdot (h_{u}^{'}\parallel h_{v}^{'} ))=LeakyReLU((a_{src}^{T}\parallel a_{dst}^{T})\cdot (h_{u}^{'}\parallel h_{v}^{'} )) \nonumber\\
&=LeakyReLU((a_{src}^{T}\cdot h_{u}^{'})+(a_{dst}^{T}\cdot h_{v}^{'}))=LeakyReLU(\theta_{u*}+\theta_{*v} )
\end{align}

In this manner, attention importance can be derived by summing two scalars, $\theta_{u*}$ and $\theta_{*v}$. 
The former signifies the attention coefficient when $u$ is the source vertex of an edge, while the latter represents the coefficient when $v$ is the target vertex. Since these values remain globally invariant for each vertex within the same semantic graph, they only need to be computed once per vertex and can be reused for subsequent edges. Additionally, for a specific target vertex, the right-hand side of Eq.~\ref{eq:coefficient_transform} is identical ($\theta_{*v}$). It is only necessary to compare the $\theta_{u*}$ values of the neighbors to discern their relative importance, obviating the need to compute the importance of neighbors slated for elimination.

\subsection{Neighbor Pruning Method Based on Min-heap}
\label{sec:pruning_method}

It is worth noting that our primary objective is to identify the top-K neighbors of a target vertex with the largest attention importance. While there is no need to sort all neighbors, the resulting top-K neighbors do not need to be ordered either. Based on these premises, we propose a neighbor pruning method inspired by the concept of min-heap\revise{,} as shown in Algorithm~\ref{alg:neighbor_prune}.

\begin{algorithm}[!ht]
    \SetAlgoLined
    \DontPrintSemicolon
    \SetKwBlock{DoParallel}{in parallel do}{end}
    \caption{\textbf{Runtime Neighbor Pruning Method}}
    \footnotesize
    \label{alg:neighbor_prune}
    \textbf{Input:} $h_u'$ of $u$ in $\mathcal{N}_v$, $h_v'$, $rd_v$, pre-trained $a^{\mathcal{P}}$ under semantic $\mathcal{P}$;\\
    \textbf{Output:} updated $rd_v$, $\alpha_{uv}$ for each retained neighbor $u$;\\
    \textbf{Initial:} predefined pruning threshold $K$. \\
        $\theta_{*v}^{\mathcal{P}}$=Compute\_Coefficient($a_{dst}^{\mathcal{P}}$, $h_v'$);\\
        \For{each neighbor $u$ in $\mathcal{N}_v$}
        {
            $\theta_{u*}^{\mathcal{P}}$=Compute\_Coefficient($a_{src}^{\mathcal{P}}$, $h_u'$); \\
            \If{$rd_v$.size $<$ $K$ }{
                push $\theta_{u*}^{\mathcal{P}}$ into $rd_v$; \Comment{\textit{$rd_v$ is not full}} \hfill\\
                \DoParallel{
                    heapify($rd_v$); \\
                    Compute\_Importance($\theta_{u*}^{\mathcal{P}}$, $\theta_{*v}^{\mathcal{P}}$);
                }
            }
            \ElseIf{$\theta_{u*}^{\mathcal{P}}$ $>$ $rd_v[0]$}
                {
                    $rd_v$[0]=$\theta_{u*}^{\mathcal{P}}$;  \Comment{\textit{replace $rd_v[0]$, discard the corresponding neighbor}}\\
                    \DoParallel{
                        heapify($rd_v$); \\
                        Compute\_Importance($\theta_{u*}^{\mathcal{P}}$, $\theta_{*v}^{\mathcal{P}}$);
                    }
                }
            \Else{
                discard $\theta_{u*}^{\mathcal{P}}$; \Comment{\textit{discard the corresponding neighbor $u$}}
            }
        }
\end{algorithm}

 A min-heap is a binary-tree-based data structure where each parent vertex has a value less than or equal to the values of its children. In a min-heap, the root vertex contains the minimum element of the whole heap. As illustrated in Algorithm~\ref{alg:neighbor_prune}, for a target vertex $v$ in semantic graph $G^\mathcal{P}$, upon computing the attention coefficient of a neighboring vertex $u$, if $v$'s retention domain (abbreviated as $rd_v$) is not at capacity, $\theta_{u*}^{\mathcal{P}}$ is directly pushed into $rd_v$, while ensuring the maintenance of the min-heap structure (lines 7-13). When $rd_v$ reaches capacity, each newly computed $\theta_{u*}^{\mathcal{P}}$ undergoes comparison with the root vertex's value ($rd_v[0]$). If the coefficient is less than $rd_v[0]$, it is inferior to all values in $rd_v$ and the corresponding neighboring vertex $u$ can be discarded instantly (line 22); the same applies when they are equal. Conversely, if the coefficient exceeds $rd_v[0]$, then $rd_v[0]$ is replaced with $\theta_{u*}^{\mathcal{P}}$, and the min-heap structure is reestablished starting from the top of the heap (lines 14-20). Following the comparison of all $\theta_{u*}^{\mathcal{P}}$ with $rd_v[0]$ and the execution of the corresponding operations, $rd_v$ contains only the attention coefficients of the neighboring vertices that have the largest attention importance for the target vertex $v$. The worst-case time complexity of this method is $O(n log K)$, where $n$ signifies the number of neighbors of a target vertex, and $K$ denotes the predefined pruning threshold.



\subsection{Parallel Execution with Operation Fusion}
\label{sec:fusion}

\begin{figure*}[!htb] 
	\centering
	\includegraphics[width=1\textwidth]{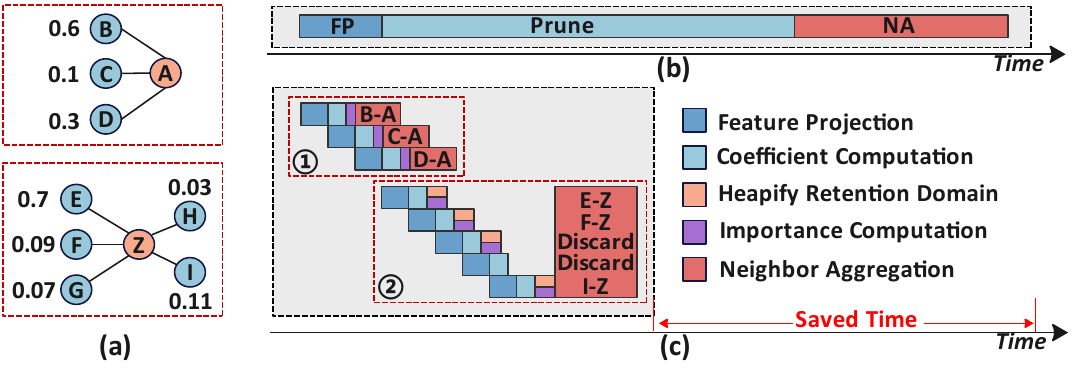}
	\caption{Illustration of operation fusion: (a) A toy graph example; (b) Staged execution with neighbor pruning; (c) Parallel execution with operation fusion.}
	\label{fig:operation_fusion}
\end{figure*}

Based on the pruning method mentioned above, we further introduce a novel HGNN execution flow with operation fusion to amortize the pruning overhead as well as harness the parallelism between execution stages. This is achieved by decoupling and reorganizing the execution stages in the original HGNN models. Fig.~\ref{fig:operation_fusion}(a) presents a toy graph with two target vertices to exemplify the operation fusion execution approach, assuming a pruning threshold $K=3$. As shown in Fig.~\ref{fig:operation_fusion}(b), traditional platforms employ a staged execution approach where the FP stage is initially executed for all vertices, followed by the centralized execution of the NA stage for aggregating neighbor information across all target vertices. Directly integrating neighbor pruning into this execution flow would introduce unacceptable time overheads as discussed in Sec.~\ref{sec:challenge}.

Our method deviates from the traditional staged execution by scheduling operations based on individual edges. For target vertices with a number of neighbors less than or equal to the pruning threshold $K$, such as vertex $A$, we first conduct FP on the source and target vertices for each incoming edge, followed by the calculation of attention coefficients and importance. Subsequently, aggregation for that edge can be directly carried out. While stages remain serial for each edge, different operations between edges are fused as in Fig.~\ref{fig:operation_fusion}(c)\blackcircledempty{1}. As for target vertex $Z$ with a number of neighbors greater than $K$, the difference in execution flow from the above process is that after obtaining the projected features, a process elaborated in Algorithm~\ref{alg:neighbor_prune} is utilized to determine whether the source vertex of an edge should be discarded, as shown in Fig.~\ref{fig:operation_fusion}(c)\blackcircledempty{2}. Neighbor vertices with attention coefficients less than the top element in the retention domain, such as vertex $H$ are directly discarded without further processing. For vertices that cannot be discarded temporarily, such as other neighbors of $Z$, the maintenance of the min-heap after pushing them into the retention domain and the calculation of attention importance are executed in parallel. Similarly, various operations between different edges are fused, making full use of the parallelism between stages. After all edges of the target vertex $Z$ have undergone the above process, aggregation is then performed on the retained neighbors $E$, $F$ and $I$, where $G$ is replaced by $I$ as depicted in line 15 of Algorithm~\ref{alg:neighbor_prune}.

\section{Architecture Design}

\subsection{Hardware Components}

\begin{figure*}[!htb] 
	\centering
	\includegraphics[width=1\textwidth]{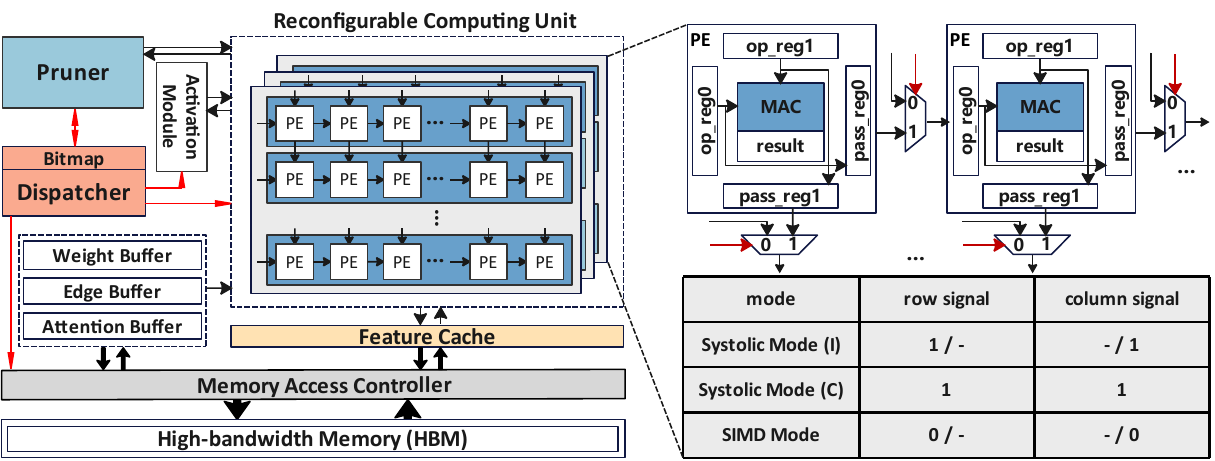}
	\caption{Overall architecture of ADE-HGNN.}
	\label{fig:architecture}
\end{figure*}

Fig.~\ref{fig:architecture} depicts the overall architecture of our proposed novel HGNN accelerator designated as ADE-HGNN. Throughout the entire execution of the HGNN models, computational workloads are mainly occupied by matrix-matrix multiplication (MM) and element-wise (EW) operations. To boost their efficiency, prior works adopt a dedicated module for each of them~\cite{HyGCN}, i.e., the systolic module and the SIMD module. Although the hybrid structure simplifies the hardware design and implementation, it leads to a significant imbalance in hardware resource utilization. To alleviate this inefficiency, we design a unified \textit{Computing Unit}, as depicted in Fig.~\ref{fig:architecture}, which can be dynamically reconfigured into systolic mode and SIMD mode. Various selector encodings enable the configuration of computing units into distinct computation modes. The independent systolic mode (I) is employed for fine-grained matrix-vector multiplication, while the combinational systolic mode (C) is utilized for MM operations. Element-wise operations are performed using SIMD mode. In addition, an \textit{Activation Module} is built to perform non-linear functions like \textit{LeakyReLU}, \textit{Elu}, and \textit{Softmax}.

The high-bandwidth memory (HBM) stores information about the original semantic graphs, mainly including the adjacent information in the compressed sparse column (CSC) format and raw features of vertices. \textit{Feature Cache} aims to cache the projected features and the intermediate neighbor aggregated features for data reuse using \textit{Least Frequently Used} strategy. \textit{Weight Buffer} and \textit{Attention Buffer} are utilized for buffering pre-trained model parameters and attention coefficients, while \textit{Edge Buffer} temporarily stores the edges which will be processed subsequently. The \textit{Memory Access Controller} is used to schedule the data interactions between the on-chip memory and HBM. To coordinate workloads across various hardware components, we introduce a centralized \textit{Dispatcher} integrated a redundancy-aware bitmap to support the reuse of intermediate results, specifically the projected features and attention coefficients. And the \textit{Pruner} is employed to determine which neighbors of each target vertex will participate in its aggregation.

\subsection{Design of Pruner}
\label{sec:pruner}

\begin{figure*}[!ht] 
	\centering
	\includegraphics[width=0.96\textwidth]{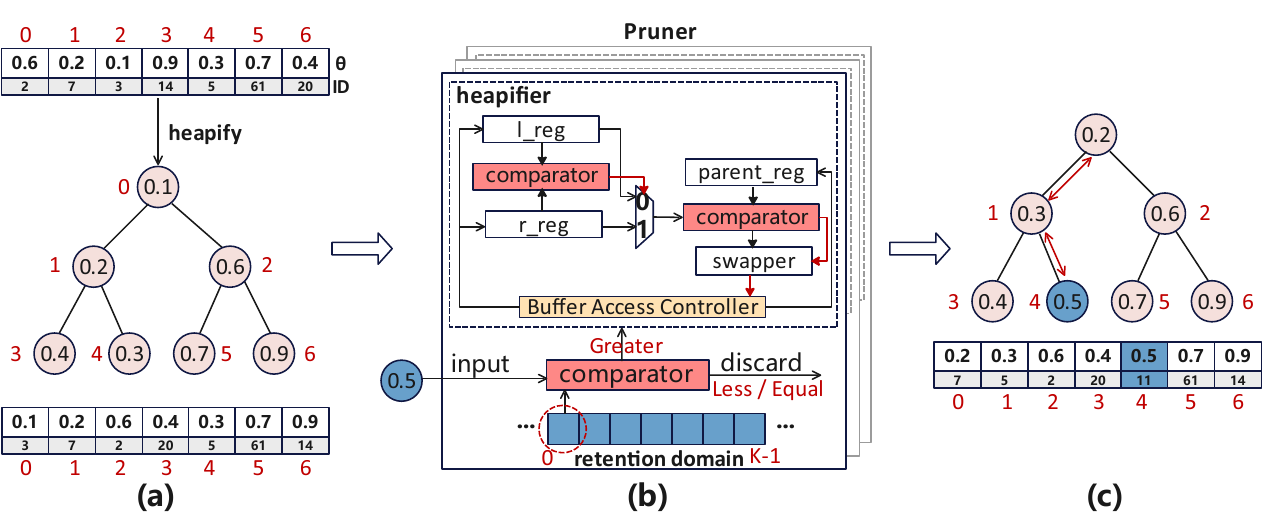}
	\caption{Pruner based on min-heap: (a) An example of a min-heap; (b) Microarchitecture of pruner; (c) Process of pushing a neighbor into retention domain.}
	\label{fig:pruner}
\end{figure*}

As depicted in Fig.~\ref{fig:pruner}(b), the pruner comprises multiple pruning units to facilitate concurrent processing of multiple target vertices. Each basic pruning unit consists of a heapifier and a contiguous on-chip storage area for the retention domain. The former organizes the data in the retention domain into a min-heap structure, where element 0 represents the minimum value (Fig.~\ref{fig:pruner}(a)). While the latter stores the attention coefficients and their corresponding information of the top-K neighboring vertices of a specific target vertex.

Each heapifer is constituted by several registers, comparators, and a swapper to achieve element comparison and exchange. The size of each pruning unit's retention domain is twice the predefined pruning threshold $K$. In addition to storing the attention coefficients of neighboring vertices, it also records the vertex IDs of the retained source vertices for subsequent neighbor aggregation. 
Since the threshold $K$ is variable, the retention domains of multiple pruning units are shared, enabling the contiguous combination of these domains to support larger $K$ values. The workflow of the pruner is elucidated in Algorithm~\ref{alg:neighbor_prune}. As shown in Fig.~\ref{fig:pruner}(c), repositioning the minimum element at the heap's top requires only $logK$ comparisons followed by exchanges and can be executed in parallel with other operations as in Sec.~\ref{sec:fusion}.


%% file: experiments.tex
\section{Experimental Results}

\subsection{Methodology}

The performance and energy consumption of ADE-HGNN are evaluated using the following tools.

\textit{\textbf{Cycle-accurate Simulator.}} 
We implement ADE-HGNN in a cycle-accurate simulator to assess its performance in terms of execution cycles, which provides detailed micro-architectural modeling for each on-chip module. Additionally, we integrate Ramulator~\cite{ramulator}, a widely-adopted cycle-accurate DRAM simulator, to precisely simulate off-chip memory accesses to HBM.

\textit{\textbf{CAD Tools.}} 
The Synopsys Design Compiler is utilized in conjunction with the TSMC 12$nm$ standard VT library for the synthesis of the RTL version of each module, while power consumption is estimated through Synopsys PrimeTime PX. The slowest module exhibits a critical path delay of 0.86 ns, putting ADE-HGNN comfortably at the 1 GHz clock frequency.

\textit{\textbf{Memory Measurements.}} 
We estimate the access latency, energy, and area of the on-chip memory components using Cacti 6.5~\cite{CACTI}. Four different scaling factors are employed to adapt these estimates to the 12nm technology node as in work~\cite{technology_scale}. The access latency and energy consumption of HBM1.0 are simulated using Ramulator and estimated at 7 pJ/bit, as reported in the literature~\cite{7pj}. 

\textit{\textbf{Benchmark.}}
We conduct experiments on three mainstream HGNN models including HAN~\cite{HAN}, RGAT~\cite{R-GAT} and SimpleHGN~\cite{Simple-HGN}, which are implemented based on the DGL 1.0.2 framework\cite{DGL} and evaluated on NVIDIA GPU T4 and the advanced GPU A100. Detailed specifications of the platforms utilized are presented in Table~\ref{tb:platform}. All models are standardized with the same hidden dimension (64) and number of attention head (8), while the layer counts are set to 1, 3, and 2, respectively. Performance metrics of GPUs are obtained using NVIDIA Nsight Compute, with all experiments conducted using data formatted in Float32. We employ three datasets (ACM, IMDB and DBLP) provided in the implementation of work~\cite{openhgnn}, where the largest dataset DBLP possesses more than 12 million edges.

\begin{table}[!ht]
\centering
\caption{Specifications of the baseline platforms and ADE-HGNN.} 
\label{tb:platform}
\resizebox{0.95\textwidth}{!}{
\begin{tabular}{ccccll}
\toprule
 & \textbf{T4} & \textbf{A100}   & \multicolumn{3}{c}{\textbf{ADE-HGNN}}                                                                                                 \\ \midrule
\textbf{\begin{tabular}[c]{@{}c@{}}Peak\\ Performance\end{tabular}}  & \begin{tabular}[c]{@{}c@{}} 8.1 TFLOPS, \\  1.59 GHz \end{tabular}                 & \begin{tabular}[c]{@{}c@{}} 19.5 TFLOPS, \\  1.41 GHz \end{tabular}                & \multicolumn{3}{c}{\begin{tabular}[c]{@{}c@{}} 16.38 TFLOPS, \\  1.0 GHz \end{tabular}} \\ \midrule
\textbf{\begin{tabular}[c]{@{}c@{}}On-chip\\ Memory\end{tabular}}  & \begin{tabular}[c]{@{}c@{}}1.28 MB (L1 Cache),\\4 MB (L2 Cache)\end{tabular} & \begin{tabular}[c]{@{}c@{}}20 MB (L1 Cache),\\40 MB (L2 Cache) \end{tabular} & \multicolumn{3}{c}{\begin{tabular}[c]{@{}c@{}}2.44 MB (Weight Buffer),\\1.20 MB (Edge Buffer),\\0.40 MB (Attention Buffer),\\ 5.00 MB (Feature Cache)\end{tabular}} \\ \midrule
\textbf{\begin{tabular}[c]{@{}c@{}}Off-chip\\ Memory\end{tabular}} & \begin{tabular}[c]{@{}c@{}} 300 GB/s, \\16 GB (GDDR6)\end{tabular}                                & \begin{tabular}[c]{@{}c@{}}2039 GB/s, \\80 GB (HBM2e)\end{tabular}                                & \multicolumn{3}{c}{\begin{tabular}[c]{@{}c@{}}512 GB/s, \\2 GB (HBM1.0)\end{tabular}}                                                                                                                            \\ \bottomrule
\end{tabular}
}
\scriptsize
\vspace{-12pt}
\end{table}

\subsection{Overall Results}

\subsubsection{Speedup}

\begin{figure*}[!hptb] 
    \centering
    \includegraphics[width=1\textwidth]{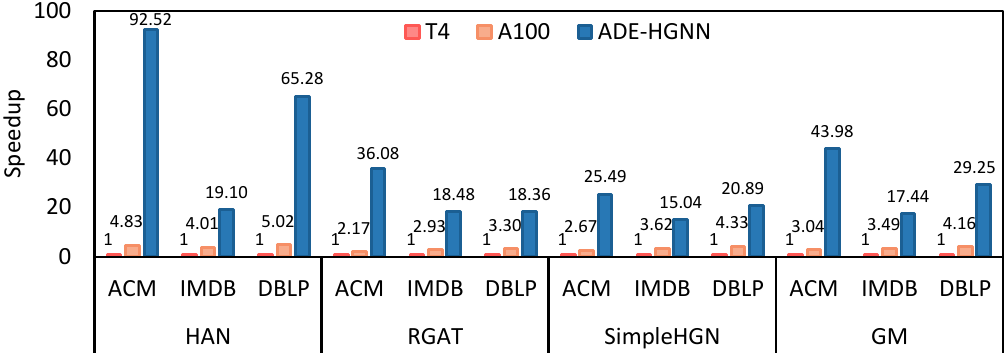}
    \caption{Speedup over GPU T4 and A100.}
    \label{fig:speedup}
\end{figure*}

As illustrated in Fig.~\ref{fig:speedup}, ADE-HGNN achieves an average speedup of 28.21$\times$ and 7.98$\times$ compared to NVIDIA GPU T4 and GPU A100, respectively. The final set of bars represents the geometric mean (GM) across all models.

The performance benefits of ADE-HGNN stem from three main factors. Initially, neighbor pruning based on attention importance drastically reduces the computational workload and off-chip memory accesses during aggregation. Furthermore, the operation-fusion-based execution flow of HGNN not only amortizes the overhead of pruning but also maximizes the parallelism between different stages. Lastly, the unified architectural design sustains a high overall utilization of computational resources, thereby further enhancing the model's execution performance.

\subsubsection{DRAM Access and Energy}
\begin{figure*}[!hptb] 
    \centering
    \includegraphics[width=0.98\textwidth]{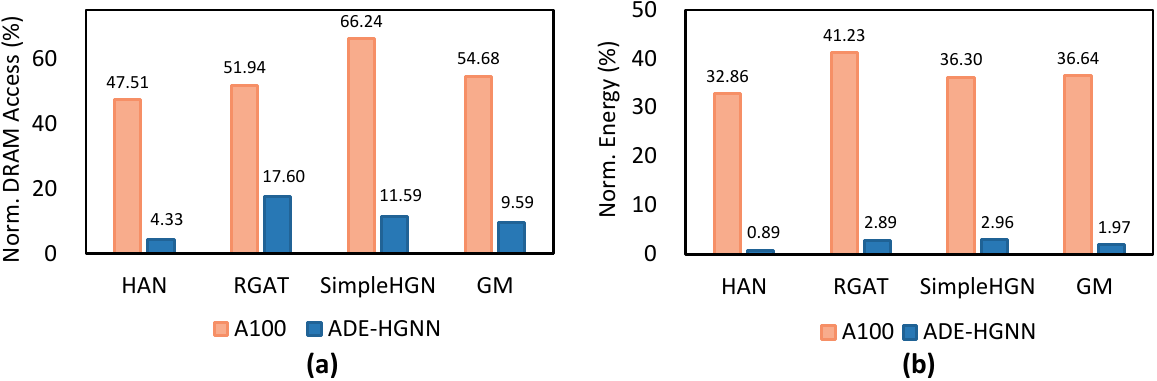}
    \caption{Detail results on DBLP datasets normalized to GPU T4: (a) DRAM access; (b) Energy consumption.}
    \label{fig:dram_energy}
\end{figure*}

As depicted in Fig.~\ref{fig:dram_energy}(a), ADE-HGNN achieves average DRAM accesses savings of 90.41\% and 82.45\% compared to GPU T4 and GPU A100, respectively. Here we only showcase the experimental outcomes obtained on the largest DBLP dataset for simplicity. First, the decomposition of attention importance computation, as illustrated in Eq.~\ref{eq:coefficient_transform}, enables the reuse of attention coefficients, eliminating a large number of redundant memory accesses. Additionally, neighbor pruning based on min-heap eliminates a substantial number of unnecessary off-chip memory accesses of unimportant neighbors. Finally, the parallel execution flow based on operation fusion enables immediate utilization of intermediate results, consequently reducing numerous unnecessary memory access operations.

Fig.~\ref{fig:dram_energy}(b) illustrates that ADE-HGNN reduces energy consumption by 98.03\% and 94.63\% compared to GPU T4 and GPU A100, respectively. Apart from the reduced off-chip memory accesses, another notable reduction in energy consumption arises from the decreased computational workload. Reusing attention coefficients within the semantic graph eliminates numerous redundant calculations. Furthermore, the neighbor pruning method diminishes the aggregation of many unimportant neighbors, resulting in a substantial reduction in computation.

\subsubsection{Area and Power}
ADE-HGNN incorporates 9.24MB of on-chip SRAM storage (including 0.20MB retention domain memory of the pruner), 8K processing elements (PEs), and other control units, with a total area of 13.68 $mm^2$ and a power consumption of 6.04 $W$. On-chip memory components collectively occupy 48.73\% of the area and consume 12.45\% of the power. The computing unit, considered the predominant module, accounts for 49.73\% of the area and 86.68\% of the power. In contrast, the pruner consisting of 128 basic pruning units occupies only 1.05\% of the area and consumes 0.27\% of the power of the entire accelerator, which is deemed negligible considering the substantial performance gains achieved through neighbor pruning.

\subsection{Effects of proposed optimizations}
\subsubsection{Effects of Neighbor Pruning}
\label{sec:accuracy_loss}

\begin{figure}[!ht] 
	\centering
	\includegraphics[width=1\textwidth]{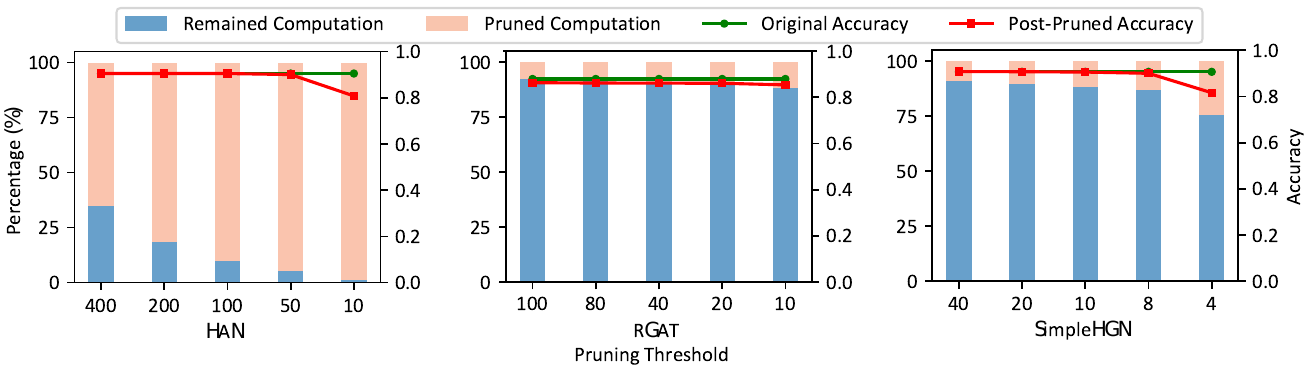}
	\caption{Pruning effect under different thresholds on ACM datasets.}
	\label{fig:pruning_effect}
\end{figure}

Fig.~\ref{fig:pruning_effect} displays the decrease in computational load and the accuracy loss resulting from pruning across three models. We only present results on the ACM dataset for simplicity, with similar characteristics observed on other datasets. As depicted in the figure, the HAN model exhibits significant redundancy in neighbor information. With the pruning threshold set to K=50 in this work, the computational load is reduced by 94.61\%, while the average accuracy loss is only 0.50\%. For the other two models, due to the smaller average degree of vertices in the datasets (relation-based SGB, while HAN utilizes metapath-based SGB), the reduction in computational load is less pronounced. However, they still achieve a reduction in computational load of 16.5\% with an accuracy loss of 0.76\%, and a reduction of 14.82\% with an average accuracy loss of 1.32\% when K=20. As the scale of real-world graph data continues to expand and the number of edges in semantic graphs increases, the method proposed in this study is anticipated to exert a more substantial impact.

\subsubsection{Effects of Operation Fusion}

Compared to the staged execution approach (without fusion) as on traditional platforms, the utilization of operation fusion enhances the performance of ADE-HGNN by an average of 3.69$\times$ across all models and datasets. These performance gains primarily stem from inter-stage parallelism. On one hand, the parallel execution between stages of HGNNs allows for the simultaneous utilization of computational and memory components. On the other hand, the fusion between the original execution stages and the neighbor pruning stages effectively amortizes almost all pruning overheads.

%% file: related_work.tex
\section{Related Work}

Given the remarkable learning capability of GNNs with graph data, GNN accelerators have garnered significant interest from the architecture community~\cite{HyGCN,FlowGNN,TopK_GAT}. Prior work\cite{TopK_GAT} utilizes neighbor pruning to achieve load balancing in GAT~\cite{GAT} model acceleration. Unfortunately, it only focuses on GAT model and fails to mitigate the substantial overhead introduced by neighbor pruning, particularly evident in HGNN models handling multiple semantic graphs with a high volume of edges. Another effort \textit{HyGCN}~\cite{HyGCN} employs two engines to accelerate the NA and FP stages, followed by inter-stage fusion to improve overall performance. However, the differing workflows of HGNNs and GNNs unavoidably result in distinct fusion methodologies.

Recent work~\cite{MetaNMP} proposes \textit{MetaNMP}, an HGNN accelerator utilizing DIMM-based near-memory processing to substantially reduce memory usage and eliminate redundant computations. \textit{MetaNMP} primarily focuses on the HGNN models utilizing intra-metapath aggregation, which hasn't been widely adopted in the algorithm community since its proposal by MAGNN~\cite{MAGNN} in 2020. Additionally, given the nascent stage of DIMM-based near-memory processing, \textit{MetaNMP} cannot be deemed a near-term solution. In contrast, our work provides a practical solution, utilizing readily available HBM and leveraging attention disparity which is more universally applied in mainstream HGNNs.

%% file: conclusion.tex
\section{Conclusion}

In order to expedite HGNN inference by exploiting attention disparity, this work introduces a runtime pruning approach based on min-heap to efficiently discard insignificant neighbors. Additionally, it presents a novel HGNN execution flow based on operation fusion to effectively amortize the pruning overhead. The newly designed HGNN accelerator, ADE-HGNN, demonstrates remarkable performance enhancements and energy efficiency gains compared to conventional GPU platforms, while maintaining negligible inference accuracy loss.

\subsubsection{\ackname} This work was supported by CAS Project for Young Scientists in Basic Research (Grant No. YSBR-029), the National Natural Science Foundation of China (Grant No. 62202451), and CAS Project for Youth Innovation Promotion Association.